\documentclass[12pt]{iopart}
\usepackage{iopams}

\begin{document}

\title{A systematic method for Schrieffer-Wolff transformation and its generalizations}

\author{Rukhsan Ul Haq$^1$ and Keshav Singh$^2$}

\address{$1$ Centre of Excellence, \\ Skoruz Technologies,  Bangalore, India}
\address{$2$ Indian Institute of Science Education and Research, Thiruvanthapuram\\
Kerala, India}
\vspace{10pt}

\begin{abstract}
Schrieffer-Wolff transformation is extensively used in quantum many-body physics to calculate the low energy effective Hamiltonian. It offers a perturbative method to understand the renormalization effects in the strong coupling regime of the quantum many-body models. The generator for Schrieffer-Wolff transformation is calculated using heuristic methods. It is highly desirable to have a rigorous, systematic and simple method for the calculation of this highly important transformation. In this paper, we put forward such method. We demontrate our method by applying it to few well-known models from various fields like strongly correlated electron systems, quantum optics and cavity electrodynamics. Our method can be applied to a broad class of quantum many-body Hamiltonians.

\end{abstract}

%
%
\submitto{\JPA}
%
\maketitle
%
%

\section{Intro}

Schrieffer-Wolff transformation(SW) was introduced in\cite{SW} to relate Anderson impurity model and Kondo model. Since then this transformation has been used very extensively in condensed matter physics. See\cite{Bravyi} as a recent reference to the literature on SW transformation. The authors have  given a mathematically rigorous treatment of the transformation. SW transformation can be understood as degenerate perturbation theory but this transformation is used to calculate the effective Hamiltonians of the models for strongly correlated electron systems and other quantum many-body systems. SW transformation can be understood as one step renormalization process which projects out the high energy excitations and leads to the low energy effective Hamiltonians which has the effects of the projected out high energy ecitations in the form of the renormalized coupling constants. Though SW transformation is perturbative in nature but it gives a method which gives physical insights about the strong coupling regime of the quantum many-body systems. One paradigmatic example is the calculation of the Kondo model as effective Hamiltonians of the Anderson imputity model. Later on, it was confirmed by numerical renormalization group calculation that Kondo model correspnds to the strong coupling fixed point of Anderson impurity model. Another important aspect of the SW transformation is its uniatry as opposed to some projective methods for calculations of the effective Hamiltonians. SW transformation is a unitary transformation which removes the off-diagonal terms to the first order and hence it is a way to diagonalization. Though this transformation is routinely used by the researchers working in condensed matter physics, it was surprising to us to find that there is no systematic method to find the generator of this transformation. In this paper we present a systematic  method to get the generator of this transformation and we will calculate the generators for five different quantum many-body Hamiltonians which are represenative models from various fields. In that way, we not only demostrate the power of our method but also its generality.

The rest of the paper is organized as follows: In the  next section, we give a brief introduction to Schrieffer-Wolff transformation and how in literature there are different approaches and interpretations. In section 3 we give the detailed description of our method. Then in the following section, we show how to apply our method to five different Hamiltonians. Finally we give our conclusions and summary of the results.

\section{What is SW transformation exactly?}
Unitary transformation is the standard method for diagonalization in Quantum Mechanics and condensed matter physics\cite{Wagner}. By the unitary transformation one changes to 
the basis in which the given Hamiltonian becomes diagonal and one achieves diagonalization in a one step process. However this is not possible for all Hamiltonains. So in latter case one tries to diagonalize the Hamiltonian in a perturbative manner. One can still use unitary transformation to achieve this goal. Schrieffer-Wolff transformation is such a method. It has been used extensively in different areas of physics under different names\cite{Bravyi}. In relativistic Quantum Mechanics  it is called Foldy-Wutheysen transformation\cite{FW},in Semiconductor physics it is called k.p perturbation theory\cite{Winkler} and in condensed matter physics it has been used as Frohlich transformation for electron-phonon problem\cite{Frohlich}.
Schrieffer-Wolff transformation not only diagonalizes the hamiltonian in a perturbative manner in which case it is unitary perturbation theory but it also renormalizes the 
parameters in the hamiltonian and hence can be thought of as a kind of renormalization procedure. SW transformation in the latter sense is used to get the effective Hamiltonian of the given Hamiltonian and hence it takes us to a particular regime of a Hamiltonain in its parameter space and it is in the sense that SW transformation was used in\cite{SW}.\\
Schrieffer-Wolff transformation is a unitary transformation. So one chooses the proper unitary operator which can either fully diagonalize the hamiltonian or to some desired order.  

\begin{equation}
H'=U^{\dagger}HU \\
H' = e^{S}H e^{-S}
\end{equation}
where S is the generator of this transformation and is an anti-hermitian operator. Usually one requires of this transformation to cancel the off-diagonal terms to the first order so that following condition is satisfied.
\begin{equation}
\left[S,H_{0}\right]=-H_{v}
\end{equation}
Expanding the operator exponential using BCH formula one gets series expansion for the transformed hamiltonian $H' $
\begin{equation}
H'=H_{0}+\frac{1}{2}\left[S,H_{v}\right]+\frac{1}{3}\left[S,\left[S,H_{v}\right]\right]+....
\end{equation}
where$H_{0}$ and $H_{v}$ are diagonal and off-diagonal parts of the hamiltonian $H$. Since the off-diagonal term gets cancelled to the first order so the effective hamiltonian to the second order is given by
\begin{equation}
H_{eff} = H_{0} + \frac{1}{2}\left [S,H_{v} \right]
\end{equation}
 
Schrieffer-Wolff transformation being very important transformation has been generalized in various ways. One very important generalization was done by Wegner\cite{Wegner} and Glazek and Wilson\cite{Glazek} independently. The new method has been called Flow equation method by Wegner and Similarity Renormalization by Glazek and Wilson. In flow equation method the unitary transformation is once again used but it is done in a continous fashion because the generator depends on the flow parameter. The relation between SW transformation and flow equation method has been worked out in \cite{Kehrein}. SW transformation has  been generalized to dissipative quantum systems as well.\cite{Kessler} \\
In  addition to method of SW transformation we have described above where we have used a generator to carry out the transformation  there is yet another method of doing the transformation in which one uses Projection operators\cite{Hewson} or Hubbard operators\cite{Fazekas}. Though using projection operator method we get the effective hamiltonian 
but it does not recover all the terms which one gets in generator method. Projection operator method is in spirit of SW transformation being a renormalization procedure. The 
Hubbard operator method is particularly suited for Hubbard model.

\section{New method for Schrieffer-Wolff transformation}
The most crucial step in doing SW transformation is to get the generator of the transformation. Once the generator is calculated the rest of the calculation is quite 
straightforward. So having an explicit method for calculating the generator is of immense value. Here We would emphasize that in the literature there is no explicit method to 
calculate the generator and hence carry out the transformation directly from the hamiltonian. In \cite{Wagner} \cite{Coleman} generator itself is obtained in a perturbative 
manner. So one does not get the generator rather one develops it in an order by order manner. In \cite{Phillips} the generator is obtained by guessing it which assumes some 
experience with the  method. In \cite{Gulasci} though the authors have summed the series to all orders but they have not described how they get the generator in the first palce. \\
In this section we  present the method for a general hamiltonian and then in the following sections we will calculate the generator of specific Hamiltonians using this explicit method.\\
  Let H be our full hamiltonian and $H_{0}$ be the diagonal part and $H_{v}$ be the off-diagonal part of the full  Hamiltonian. To obtain the generator we will proceed in two 
steps. In the first step we will find the commutator $[H_{0},H_{v}]$ and call it $\eta$. In the second step we will impose the condition of removing the off-diagonal part till
 first order on $\eta$. To do that we will have to keep the coefficients undetermined and they will be determined by the  above condition. So $\eta$ has to statisfy $[\eta,H_{0}]= -H_{v}$ in order to be the generator of the transformation. The latter condition determines the coefficients and we get the generator for SW transformation of the  given 
hamiltonian. In the next section we will calculate the generator of SW transformation for Single Impurity  Anderson Model for which the transformation was carried out in the 
original paper\cite{SW}.

\section{Generator for Anderson Impurity Model}
 We will first write down the single impurity Anderson Hamitonian in second quantized notation:
\begin{equation} 
H = \sum_{k\sigma}\epsilon_{k}c^{\dag}_{k\sigma}c_{k\sigma}+\sum_{\sigma}\epsilon_{d}d^{\dag}_{\sigma}d_{\sigma}+\sum_{k\sigma}V_{k}(c^{\dag}_{k\sigma}d_{\sigma}+d^{\dag}_{\sigma}c_{k\sigma})+ Un_{d\uparrow}n_{d{\downarrow}}
\end{equation}
The hamiltonian has one off-diagonal term which we call as $H_{v}$ and diagonal terms which together we call $H_{0}$.
\begin{equation}
H_{0}=\sum_{k\sigma}\epsilon_{k}c^{\dag}_{k\sigma}c_{k\sigma}+\sum_{\sigma}\epsilon_{d}d^{\dag}_{\sigma}d_{\sigma}+ Un_{d\uparrow}n_{d{\downarrow}} 
\end{equation}
\begin{equation}
H_{v}=\sum_{k\sigma}V_{k}(c^{\dag}_{k\sigma}d_{\sigma}+d^{\dag}_{\sigma}c_{k\sigma})
\end{equation}
Now the first step is to calculate $\eta$ which is basically commutator of diagonal part with off-diagonal part of the hamiltonian.

Now the first step is to calculate $\eta$ which is basically commutator of diagonal part with off-diagonal part of the hamiltonian.
\begin{eqnarray}
\eta & = \left [H_{0},H_{v}\right ] \\
\eta & =\left [\sum_{k\sigma}\epsilon_{k}c^{\dag}_{k\sigma}c_{k\sigma}+\sum_{\sigma}\epsilon_{d}d^{\dag}_{\sigma}d_{\sigma}+
Un_{d\uparrow}n_{d{\downarrow}},\sum_{k\sigma}V_{k}(c^{\dag}_{k\sigma}d_{\sigma}+d^{\dag}_{\sigma}c_{k\sigma})\right] \\
\eta & = \sum_{k\sigma}(\epsilon_{k}-\epsilon_{d}-Un_{d\bar{\sigma}})V_{k}(c^{\dag}_{k\sigma}d_{\sigma}-d^{\dag}_{\sigma}c_{k\sigma})
\end{eqnarray}
In the second step we will impose the condition of removing the off-diagonal term to the first order. To do that we will keep the
coefficients undetermined and actually they will get determined automatically once $\eta$ satisfies the condition. We will label it with S  to emphasize that it is not actually $\eta$ which is the generator rather it is S with correct coefficients. What $\eta$ has similar to S is the form of the operators.
\begin{equation}
S =\sum_{k\sigma}(A_{k}-B_{k}n_{d\bar{\sigma}})V_{k}(c^{\dag}_{k\sigma}d_{\sigma}-d^{\dag}_{\sigma}c_{k\sigma})
\end{equation}
Now we will impose te condition on S to determine $A_{k}$ and $ B_{k}$
\begin{eqnarray}
&\left [S,H_{0}\right ]   =  - H_{v} \\
&\Rightarrow \left [ \sum_{k\sigma} A_{k}(\epsilon_{d}-\epsilon_{k}) +
  \sum_{k\sigma}(A_{k}U-B_{k}(\epsilon_{d}-\epsilon_{k}+U)n_{d\bar{\sigma}})\right ] \\ 
&  (V_{k}(c^{\dag}_{k\sigma}d_{\sigma}+  d^{\dag}_{\sigma}c_{k\sigma}) \\
& =  -\sum_{k\sigma} V_{k}(c^{\dag}_{k\sigma}d_{\sigma} + d_{\sigma}c^{\dag}_{k\sigma}) \\
&\Rightarrow A_{k}(\epsilon_{d}-\epsilon_{k}) + (A_{k}U+ B_{k}(\epsilon_{d}-\epsilon_{k}+U)n_{d\bar{\sigma}} =  -1 
\end{eqnarray}
Solving for $A_{k}$ and $B_{k}$ we obtain:
\begin{eqnarray}
A_{k}& =\frac{1}{\epsilon_{k}-\epsilon_{d}}\\
B_{k}& = \frac{1}{\epsilon_{k}-\epsilon_{d}-U} - \frac{1}{\epsilon_{k}-\epsilon_{d}}
\end{eqnarray}
In this way we have calculated the generator of SW transformation for Single Impurity Anderson Model.
\begin{equation}
S = \sum_{k\sigma}(A_{k} + B_{k}n_{d\bar{\sigma}})V_{k}(c^{\dag}_{k\sigma}d_{\sigma}-d^{\dag}_{\sigma}c_{k\sigma}) 
\end{equation}
\begin{eqnarray}
A_{k} & = \frac{1}{\epsilon_{k}-\epsilon_{d}} \\ 
B_{k} & = \frac{1}{\epsilon_{k}-\epsilon_{d}-U} - \frac{1}{\epsilon_{k}-\epsilon_{d}}
\end{eqnarray}
We can write the generator in the same form as was written in\cite{SW} by using extra index $\alpha$ which takes two values.
\begin{equation}
\mathrm{S} =\sum_{k\sigma\alpha}\frac{V_{k}}{\epsilon_{k}-\epsilon_{\alpha}}n^{\alpha}_{d,\bar{\sigma}}c^{\dag}_{k\sigma}d_{\sigma}-h.c.
\end{equation} 
\begin{eqnarray}
n_{d\bar{\sigma}}^{\alpha} & = n_{d\bar{\sigma}} & \epsilon_{\alpha}=\epsilon_{d}+U \quad \alpha = + \\
&=1-n_{d\bar{\sigma}}    &\epsilon_{\alpha}=\epsilon_{d}      \quad  \alpha = -
\end{eqnarray}
Summing over $\alpha$ we get $\mathrm{S}$ in the same form as we have calculated.
\begin{eqnarray}
\mathrm{S} & = \sum_{k\sigma}[\frac{V_{k}}{\epsilon_{k}-\epsilon_{d}-U} n_{d\bar{\sigma}}c^{\dag}_{k\sigma}d_{\sigma}+ \\
& \frac{V_{k}}{\epsilon_{k}-\epsilon_{d}}(1-n_{d\bar{\sigma}})c^{\dag}_{k\sigma}d_{\sigma}] - h.c.
\end{eqnarray}
Rearranging the terms one gets $\mathrm{S}$ in the form as we have calculated above.

\section{Generator for Periodic Anderson Model}
Periodic Anderson Model(PAM) is the standard model in heavy fermion physics. We will write down the model in standard second quantized notation as given in \cite{Nolting}
\begin{equation}
H = \sum_{k\sigma}\epsilon_{k}n_{k\sigma}+\sum_{i\sigma}\epsilon_{f}n^{f}_{i\sigma}+U\sum_{i}n_{i\uparrow}n_{i\downarrow}+
\sum_{ki\sigma}(V_{k}e^{-ikR_{i}}c^{\dag}_{k\sigma}f_{i\sigma}+h.c)
\end{equation}
To calculate the generator of SW transformation for PAM we will proceed as per the method given in section III. So we will first calculate $\eta$ as follows:
\begin{eqnarray}
\eta & = [H_{0},H_{v}] \\
&\eta =\left [\sum_{k\sigma}\epsilon_{k}n_{k\sigma}+\sum_{i\sigma}\epsilon_{f}n^{f}_{i\sigma}+U\sum_{i}n_{i\uparrow}n_{i\downarrow},\sum_{ki\sigma}(V_{k}e^{-ikR_{i}}c^{\dag}_{k\sigma}f_{i\sigma}+h.c)\right ] \\
&\eta = \sum_{ki\sigma}(\epsilon_{k}-\epsilon_{f}-Un_{i\bar{\sigma}})V_{k}(c^{\dag}_{k\sigma}f_{i\sigma}e^{-ikR_{i}} - h.c.)
\end{eqnarray}
So we can write the generator for PAM as:
\begin{equation}
S = \sum_{ki\sigma}(A_{k} + B_{k}n_{i\bar{\sigma}})V_{k}(c^{\dag}_{k\sigma}f_{i\sigma}e^{-ikR_{i}}-h.c)
\end{equation}
Where $A_{k}$ and $B_{k}$ need to be determined. To do so we will follow the second step of the method and proceed as follows.
\begin{eqnarray}
&\left[S,H_{0}\right] =  - H_{v} \\
 &\bigg[\sum_{ki\sigma}(A_{k}+ B_{k}n_{i\bar{\sigma}})V_{k}(c^{\dag}_{k\sigma}f_{i\sigma}e^{-ikR_{i}}-h.c), \\
&\sum_{k'\sigma'}\epsilon_{k'}n_{k'\sigma'}+
\sum_{i\sigma'}\epsilon_{f} n_{i\sigma'}+U\sum_{i}n_{i\uparrow}n_{i\downarrow} \bigg] \\
& = - H_{v}\\ 
&\Rightarrow \sum_{ki\sigma}(A_{k}(\epsilon_{f}-\epsilon_{k}+Un_{i\bar{\sigma}})+B_{k}n_{i\bar{\sigma}}(\epsilon_{f}-\epsilon_{k}+U) \\
&V_{k}(c^{\dag}_{k\sigma}f_{i\sigma}e^{-ikR_{i}} + h.c.))\\
&= - V_{K}  (c^{\dag}_{k\sigma}f_{i\sigma}e^{-ikR_{i}} +h.c.)\\
&\Rightarrow A_{k}(\epsilon_{f}-\epsilon_{k})+( A_{k}U +B_{k}(\epsilon_{f}-\epsilon_{k}+U))n_{i\bar{\sigma}}= -1 
\end{eqnarray}
Solving for $A_{k}$ and $B_{k}$ in the above equation we get:
\begin{eqnarray}
 A_{k}& = \frac{1}{\epsilon_{k}-\epsilon_{f}} \\
 B_{k}& = \frac{1}{\epsilon_{k}-\epsilon_{f}-U} - \frac{1}{\epsilon_{k}-\epsilon_{f}}
\end{eqnarray}
This way we have got the generator of SW transformation for PAM and can be written as : 
\begin{equation}
S= \sum_{k\sigma i}(A_{k}+B_{k}n_{i\bar{\sigma}})V_{k}(c^{\dag}_{k\sigma}f_{i\sigma}e^{-ikR_{i}}-f^{\dag}_{i\sigma}c_{k\sigma}e^{ikR_{i}})
\end{equation}
where $A_{k}$ and $B_{k}$ are given by 
\begin{eqnarray}
A_{k} & = \frac{1}{\epsilon_{k}-\epsilon_{f}}\\
B_{k} & = \frac{1}{\epsilon_{k}-\epsilon_{f}-U}-\frac{1}{\epsilon_{k}-\epsilon_{f}}
\end{eqnarray}

\section{Generator for Anderson Holstein Model}
In this section, we will apply our method to calculate the Schrieffer-Wolff generator for Anderson-Holstein model which can be considered as an generalization of Anderson impurity model and it takes the effect of electron-phonon coupling into consideration. The Hamiltonian for Anderson-Holstein Model in the second quantized notation is given as:

\begin{eqnarray}
H=\sum_{k\sigma}\epsilon_{k}c_{k\sigma}^{\dagger}c_{k\sigma} 
+ \sum_{k\sigma}V_{k}(c_{k\sigma}^{\dagger}d_{\sigma} + d_{\sigma}^{\dagger}c_{k\sigma}) + \sum_{\sigma}\epsilon_{d}n_{d\sigma} 
    + Un_{d\uparrow}n_{d\downarrow}+  \\
  w_{0}b^{\dagger}b + \lambda(b^{\dagger} + b)(n_{d\uparrow} + n_{d\downarrow})
\end{eqnarray}

The Hamiltonian has two off-diagonal terms and four diagonal terms which we call as $H_{\nu}$ and $H_{0}$ respectively.
\begin{equation}
  H_{0} = \sum_{k\sigma}\epsilon_{k}c_{k\sigma}^{\dagger}c_{k\sigma} + \sum_{\sigma}\epsilon_{d}n_{d\sigma} + Un_{d\uparrow}n_{d\downarrow}
    + w_{0}b^{\dagger}b
\end{equation}
\begin{equation}
    H_{\nu} = \sum_{k\sigma}V_{k}(c_{k\sigma}^{\dagger}d_{\sigma} + d_{\sigma}^{\dagger}c_{k\sigma}) + \lambda(b^{\dagger} + b)(n_{d\uparrow} + n_{d\downarrow})
\end{equation}

Now we will calculate $\eta$ which is the commutator of diagonal part with off-diagonal part of the Hamiltonian.

\begin{eqnarray}
\eta = [H_{0},H_{\nu}] \\
\eta = \sum_{k\sigma}(\epsilon_{k} + \epsilon_{d} - Un_{d\bar{\sigma}})V_{k}(c_{k\sigma}^{\dagger}d_{\sigma} - d_{\sigma}^{\dagger}c_{k\sigma}) + w_{0}\lambda(b^{\dagger} - b)(n_{d\uparrow} + n_{d\downarrow})
\end{eqnarray}
As we have seen earlier,  $\eta$ gives the operator form of the generator S.

\begin{equation}
S = \sum_{k\sigma}(A_{k} - B_{k}n_{d\bar{\sigma}})V_{k}(c_{k\sigma}^{\dagger}d_{\sigma} - d_{\sigma}^{\dagger}c_{k\sigma}) + C(b^{\dagger} - b)(n_{d\uparrow} + n_{d\downarrow})
\end{equation}

The undetermined coefficients $A_{k}$, $B_{k}$ and C can be determined by imposing the condition on S, to remove the off-diagonal terms $H_{v}$  up to the first order.
\begin{equation}
    [S,H_{0}] = -H_{\nu}
\end{equation}

\begin{eqnarray}
\bigg[\sum_{k\sigma}(A_{k} - B_{k}n_{d\bar{\sigma}})V_{k}(c_{k\sigma}^{\dagger}d_{\sigma} - d_{\sigma}^{\dagger}c_{k\sigma}) + C(b^{\dagger} - b)(n_{d\uparrow} + n_{d\downarrow}),\\
    \sum_{k\sigma}\epsilon_{k}c_{k\sigma}^{\dagger}c_{k\sigma} +
 \sum_{\sigma}\epsilon_{d}n_{d\sigma} 
 + U n_{d\uparrow} n_{d\downarrow}
   + w_{0}b^{\dagger}b\bigg] = \\
-  \sum_{k\sigma}V_{k}(c_{k\sigma}^{\dagger}d_{\sigma} + d_{\sigma}^{\dagger}c_{k\sigma}) + \lambda(b^{\dagger} + b)(n_{d\uparrow} + n_{d\downarrow})
\end{eqnarray}

Solving this equation leads to following two equations for the undetermined co-efficients:

\begin{equation}
  A_{k}(\epsilon_{d} - \epsilon{k}) + (A_{k}U + B_{k}(\epsilon_{d} - \epsilon_{k} + U))n_{d\bar{\sigma}} = -1 
\end{equation}

\begin{equation}
    -Cw_{0} = -\lambda
\end{equation}

Solving for $A_{k}$, $B_{k}$ and C we obtain:
\begin{equation}
 A_{k} = \frac{1}{\epsilon_{k} - \epsilon_{d}}
\end{equation}
\begin{equation}
    B_{k} = \frac{1}{\epsilon_{k} - \epsilon_{d} - U} - \frac{1}{\epsilon_{k} - \epsilon_{d}}
\end{equation}
\begin{equation}
    C = \frac{\lambda}{w_{0}}
\end{equation}

\section{Generator for Frohlich Hamiltonian}
In this section, we will apply our method to calculate the Schrieffer-Wolff generator for Frohlich Hamiltonian which is a well-known model studied in the area of superconductivity. The Frohlich Hamiltonian in second quantized notation is given as:
\\
\begin{eqnarray}
 H = \sum_{k}\epsilon_{k}c_{k}^{\dagger}c_{k} + \sum_{q}w_{q}(b_{q}^{\dagger}b_{q} + \frac{1}{2}) +\sum_{k,q}g_{k,q}c_{k+q}^{\dagger}c_{k}(b_{q} + b_{-q}^{\dagger})
\end{eqnarray}

We can list out the diagonal and off diagonal part of the Hamiltonian as $H_{\nu}$ and $H_{0}$ respectively:
\begin{equation}
 H_{0} = \sum_{k}\epsilon_{k}c_{k}^{\dagger}c_{k} + \sum_{q}w_{q}(b_{q}^{\dagger}b_{q}+\frac{1}{2})
\end{equation}

\begin{equation}
 H_{\nu} = \sum_{k,q}g_{k,q}c_{k+q}^{\dagger}c_{k}(b_{q} + b_{-q}^{\dagger})
\end{equation}

As the first step we calculate $\eta$, which the commutator of diagonal part with the off-diagonal part of the Hamiltonian.
\begin{equation}
  \eta = [H_{0},H_{\nu}]
\end{equation}

\begin{eqnarray}
 \eta = \bigg[\sum_{k}\epsilon_{k} c_{k}^{\dagger} c_{k} + \sum_{q}w_{q}(b_{q}^{\dagger}b_{q} + \frac{1}{2}), \sum_{kq}g_{kq}c_{k+q}^{\dagger}c_{k}(b_{q} + b_{-q}^{\dagger})\bigg]
\end{eqnarray}

\begin{equation}
    \eta = \sum_{k,q}g_{k,q}c_{k+q}^{\dagger}c_{k}[(\epsilon_{k+q} + \epsilon_{k} - w_{q})b_{q} + (\epsilon_{k+q} + \epsilon_{k} + w_{-q})b_{-q}^{\dagger}]
\end{equation}
So, we have obtained the operator form of the desired generator. 
\begin{equation}
    S = \sum_{k,q}c_{k+q}^{\dagger}c_{k}(A_{k,q}b_{q} + B_{k,q}b_{-q}^{\dagger})
\end{equation}
The undetermined coefficients $A_{k,q}$ and $B_{k,q}$ can be determined by imposing the condition on S to remove the off-diagonal terms up to first order, when the transformation is done.
\begin{equation}
    [S,H_{0}] = -H_{\nu}
\end{equation}

\begin{eqnarray}
 \bigg[\sum_{k,q}c_{k+q}^{\dagger}c_{k}(A_{k,q}b_{q} + B_{k,q}b_{-q}^{\dagger}),\sum_{k}\epsilon_{k}c_{k}^{\dagger}c_{k} + \sum_{q}w_{q}(b_{q}^{\dagger}b_{q} + \frac{1}{2})\bigg] =\\
 -[\sum_{k,q}g_{k,q}c_{k+q}^{\dagger}c_{k}(b_{q} + b_{-q}^{\dagger})]
\end{eqnarray}
\\
This leads us to two equations:
\begin{equation}
    A_{k,q}(\epsilon_{k} - \epsilon_{k+q} + w_{q}) = -g_{k,q}
\end{equation}
\begin{equation}
    B_{k,q}(\epsilon_{k} - \epsilon_{k+q} - w_{-q}) = -g_{k,q}
\end{equation}
Solving for $A_{k,q}$ and $B_{k,q}$ we obtain:
\begin{equation}
    A_{k,q} = \frac{-g_{k,q}}{\epsilon_{k} - \epsilon_{k+q} + w_{q}}
\end{equation}
\begin{equation}
    B_{k,q} = \frac{-g_{k,q}}{\epsilon_{k} - \epsilon_{k+q} + w_{-q}}
\end{equation}

\section{Generator for Jaynes-Cummings Model}
Jyanes-Cummings model is one of the paradigmatic models in the field of quantum optics. In this section, we will show that our method can be applied to carry out Schrieffer-Wolff transformation for this model also. We will derive the generator of SW transformation for Jaynes-Cummings model which can then used to obtain the effective Hamiltonian of this model. First we write Jaynes-Cummings model  in second quantized notation:
\\
\begin{eqnarray}
  H = w_{r}a^{\dagger}a - \frac{w_{q}}{2}\sigma^{z} + g(a^{\dagger}\sigma^{-} + a\sigma^{+})
\end{eqnarray}
\\
The $\sigma^{+}$ and $\sigma^{-}$ are defined as:
\begin{equation}
    \sigma^{\pm} = \frac{1}{\sqrt{2}}[\sigma^{+} \pm \sigma^{-}]
\end{equation}
We can list out the diagonal and off-diagonal parts of the Hamiltonian as $H_{0}$ and $H_{\nu}$ respectively:
\begin{equation}
    H_{0} = w_{r}a^{\dagger}a - \frac{w_{q}}{2}\sigma^{z} 
\end{equation}

\begin{equation}
    H_{\nu} = g(a^{\dagger}\sigma^{-} + a\sigma^{+})
\end{equation}
We begin by calculating $\eta$, which is the commutator of diagonal part with off-diagonal part of Hamiltonian.

\begin{eqnarray}
  \eta = \bigg[w_{r}a^{\dagger}a - \frac{w_{q}}{2}\sigma^{z},g(a^{\dagger}\sigma^{-} + a\sigma^{+})\bigg]    
\end{eqnarray}
\\

\begin{equation}
    \eta = g(w_{r} + w_{q})a^{\dagger}\sigma^{-} + g(w_{q} - w_{r})a\sigma^{+}
\end{equation}

In form of $\eta$ we have the general operator form for the generator, which we label as S.
\begin{equation}
    S = Aa^{\dagger}\sigma^{-} + Ba\sigma^{+}
\end{equation}
The undetermined coefficients A and B can be determined by imposing the condition on S to remove the off-diagonal term of the Hamiltonian up to first order, when the transformation is done.
\begin{equation}
    [S,H_{0}] = -H_{\nu}
\end{equation}
\begin{eqnarray}
    \bigg[Aa^{\dagger}\sigma^{-} + Ba\sigma^{+},w_{r}a^{\dagger}a - \frac{w_{q}}{2}\sigma^{z}\bigg] =  -[g(a^{\dagger}\sigma^{-} + a\sigma^{+})]
\end{eqnarray}
This leads us to two equations:
\begin{equation}
    B(w_{r} + w_{q}) = -g 
\end{equation}
\begin{equation}
    -A(w_{r} + w_{q}) = -g
\end{equation}
Thus we obtain the coefficients as:
\begin{equation}
    A = -B = \frac{g}{w_{r} + w_{q}}
\end{equation}

\section{Conclusions}
Schrieffer-Wolff(SW) transformation is a very important transformation in quantum many-body physics where it is routinely used to get the low energy effective Hamiltonian of  quantum many-body Hamiltonians which are difficult to be dealt analytically. Historically, SW transformation gave  very important insight  about the relation between two different Hamiltonians, Anderson impurity model and Kondo model. This transformation showed that Kondo model describes the strong coupling physics of the Anderson impurity model. Similarly, SW transformation was also used to understand the electron-phonon problem and how it gives rise to attractive effective interaction in BCS superconductivity. Given the significance of SW transformation, a systematic method for the calculation of the generator of this transformation should be considered a very important contribution. That is what we have carried out in this paper by putting forward a very systematic method which can be used to carry out SW transformation for a broad class of quantum many-body Hamiltonians across various fields from strongly correlated electron systems, nanophysics, quantum optics to cavity quantum electrodynamics. We have already demonstarted the power of this method by calculating the generator for five representative models which are celebarted models in various fields.  Our method has also been used by two different research groups for calculation of the generator of SWT to study non-Fermi liquid\cite{Kim} behavior and physics of NV centres in quantum optics\cite{Ali}. We leave to future to find many more applications of our method and we also hope that other researchers will find our method very useful in their research in quantum many-body physics. 

\section{Acknowledgements}
 Rukhsan Ul Haq would like to thank Professor N. S. Vidhyadhiraja for various discussions related to this work. Keshav would  like to thank Department of Science and Technology(DST GOI) for the Inspire Fellowship.

\end{document}